\documentclass[trackchanges, onecolumn]{aastex631}

\usepackage{wrapfig}
\usepackage{rotating}
\usepackage{hyperref}
\shorttitle{T CrB Super-remnant}
\shortauthors{Shara, Lanzetta, Masegian et al.}

\begin{document}

%%
%\watermark{DRAFT}
%%%%%%%%%%%%%%%%%%%%%%%%%%%%%%%%%%%%%%%%%%

\title[T CrB nova Super-Remnant]{The Newly Discovered Nova Super-Remnant Surrounding Recurrent Nova T Coronae Borealis: Will it Light Up During the Coming Eruption?}

\correspondingauthor{Michael Shara  mshara@amnh.org}

\author[0000-0003-0155-2539]{Michael M. Shara}
\affiliation{Department of Astrophysics, American Museum of Natural History, New York, NY 10024, USA}

\author[0000-0001-6906-7594]{Kenneth M. Lanzetta}
\affiliation{Department of Physics and Astronomy, Stony Brook University, Stony Brook, NY 11794-3800, USA}

\author[0000-0002-3361-2893]{Alexandra Masegian}
\affiliation{Department of Astronomy, Columbia University, New York, NY 10027, USA}

\author[0000-0003-2922-1416]{James T. Garland}
\affiliation{Department of Astrophysics, American Museum of Natural History, New York, NY 10024, USA}
\affil{David Dunlap Department of Astronomy, University of Toronto, 50 George Street, Toronto, ON M5S 3H4, Canada}

\author[0009-0006-3617-1356]{Stefan Gromoll}
\affiliation{Amazon Web Services, 410 Terry Ave. N, Seattle, WA 98109, USA}

\author[0000-0003-3457-0020]{Joanna Mikolajewska}
\affil{N. Copernicus Astronomical Center, Polish Academy of Sciences, Bartycka 18, 00–716 Warsaw, Poland}

\author[0000-0001-5094-3785]{Mikita Misiura}
\affiliation{Bain \& Company, 131 Dartmouth Street, Boston MA 01116, USA}

\author[0000-0002-9821-2911]{David Valls-Gabaud}
\affil{Observatoire de Paris, LERMA, CNRS UMR 8112, 61 Avenue de l’Observatoire, 75014 Paris, France}

\author[0000-0001-7796-1756]{Frederick M. Walter}
\affil{Department of Physics and Astronomy, Stony Brook University, Stony Brook, NY 11794-3800, USA}

\author[0000-0002-0004-9360]{John K. Webb}
\affil{Institute of Astronomy, University of Cambridge, Madingley Road, Cambridge CB3 0HA, United Kingdom}

\begin{abstract}
A century or less separates the thermonuclear-powered eruptions of recurrent novae in the hydrogen-rich envelopes of massive white dwarfs. The colliding ejecta of successive recurrent nova events are predicted to always generate very large (tens of parsecs) super-remnants; only two examples are currently known. T CrB offers an excellent opportunity to test this prediction. As it will almost certainly undergo its next, once-in $\sim$80-year recurrent nova event between 2024 and 2026, we carried out very deep narrowband and continuum imaging to search for the predicted, piled-up ejecta of the past millenia. While nothing is detected in continuum or narrowband [OIII] images, a $\sim$ 30-parsec-diameter, faint nebulosity surrounding T CrB is clearly present in deep H$\alpha$, [NII] and [SII] narrowband Condor Array Telescope imagery. We predict that these newly detected nebulosities, as well as the recent ejecta that have not yet reached the super-remnant, are far too optically-thin to capture all but a tiny fraction of the photons emitted by RN flashes. We thus predict that fluorescent light echoes will {\it not} be detectable following the imminent nova flash of T CrB. Dust may be released by the T CrB red giant wind in pre-eruption outbursts, but we have no reliable estimates of its quantity or geometrical distribution. While we cannot predict the morphology or intensity of dust-induced continuum light echoes following the coming flash, we encourage multi-epoch {\it Hubble Space Telescope} optical imaging as well as {\it James Webb Space Telescope} infrared imaging of T CrB during the year after it erupts.
\end{abstract}


\section{Introduction} \label{sec:intro}

\subsection{Novae}
Nova eruptions occur when an envelope of accreted, hydrogen-rich matter on the surface of a white dwarf (WD) star undergoes a thermonuclear runaway \citep{starrfield1972,prialnik1979,Yaron2005}. The binary companion that donates the hydrogen can be a red dwarf, subgiant, or giant star \citep{Warner1995}. Near peak brightness, novae shine with luminosities of $10^{4}$ to $10^{6}L_{\odot}$ for days to years, before extinguishing themselves by ejecting the accreted envelope \citep{prialnik1978}. All novae recur thousands of times \citep{Ford1978}, with inter-eruption periods typically ranging from millenia to Myr over their multi-Gyr lifetimes \citep{Hillman2020}. 

\subsection{Recurrent Novae}
The shortest inter-eruption periods define the Recurrent Nova (RN) subclass, whose members (by definition) undergo eruptions separated by less than 100 years \citep{Schaefer2010}. The two dominant parameters that control the frequency with which a nova undergoes an eruption are the WD mass, and the rate at which matter is accreted onto it \citep{Yaron2005}. The high surface gravities of massive WDs ($\gtrapprox$$1.3 M_{\odot}$) enable thermonuclear runaways to occur in very low mass envelopes \citep{Hillman2016}. If the donors are subgiants or giants, their very high rates of mass transfer ($10^{-6} - 10^{-7}M_{\odot}$/yr) \citep{walder2008,Wolf2013,booth2016} can build critical-mass envelopes, and hence RN, in under one century. 

\subsection{Nova Super-remnants}
Because novae eject matter with a range of speeds, the most-rapidly moving parts of envelopes ejected in successive RN events are predicted to overtake and collide with the slowest ejecta of the previous eruption, leading to the buildup of vast ($\sim$10 to 100 parsec) nova super-remnants (NSR) which should surround all RN \citep{Healy-Kalesh2023}. These NSR must have swept up many thousands of times more ISM mass than was ejected by the white dwarf in all its nova eruptions, so that NSR expansion velocities should be only a few tens of km/s. Remarkably, only two examples of NSR have been detected and imaged \citep{Darnley2019,Shara2024,MHK2024}. The KT Eri NSR strongly supports both the size and slow expansion speed predicted for NSR from numerical simulations \citep{Darnley2019, Healy-Kalesh2023}. 

A recent survey yielded zero new NSR surrounding 20 known RN in the Local Group of galaxies \citep{Healy-Kalesh2024b}. The extremely low predicted surface brightnesses of NSR probably explains the lack of success of this recent survey \citep{Healy-Kalesh2023}.

\subsection{T CrB}
The well-studied Galactic RN T CrB \citep{herbig1946,kraft1958,webbink1976,kenyon1986,selvelli1992,AM1999,zam2004,luna2018,mas2023} underwent well-documented nova eruptions in the years 1866 and 1946, and perhaps 1787 and 1217 \citep{brad2023}. It reached apparent magnitude at least $\sim$ 2.5 in 1866 \citep{petit1946}, and 1.7 in 1946 \citep{Shears2024}. Every RN system's eruption light curves are very similar to each other \citep{Schaefer2010}, so we can expect the peak visual brightness of T CrB to reach or slightly exceed 1.7 mag during the coming eruption. The underlying binary has an orbital period of $\sim$ 227.6 d \citep{Fekel2000}, and is comprised of a massive WD ($\sim$ 1.37 $M_{\odot}$) and a $\sim$ 1.12 $M_{\odot}$ red giant \citep{bel1998,stanishev2004}.

In addition to optical emission, T CrB emits from the radio \citep{linford2019} through the infrared \citep{mas2023}, ultraviolet \citep{luna2018}, and x-ray \citep{luna2019} portions of the electromagnetic spectrum. Pre-outburst observations of T CrB were carried out by ALMA on 10 August 2024 at 107.9, 136.4 and 359.5 Ghz (ATel 16781). They reveal a faint point source at the position of T CrB detected with at least 4 sigma significance, consistent with a power-law with an index of approximately +0.6. High resolution spectra at the locations of CO, SiO and HCN lines revealed no line emission. 

Strong optical emission lines (roughly 12 {\AA} wide, implying $\sim$ 3,000 km/s velocity ejection) were observed in 1946 and 1866 \citep{mcl1946}. Thus we know that there is pre-existing material surrounding T CrB, and still expanding away from it, that will be a ``target" for the ejecta and the photons of the upcoming and future eruptions. Locating that pre-existing material now, {\it before the next eruption}, would allow us to {\it predict} the locations of at least some of the light echoes \citep{Falla2003,shara2015} that previous generations of T CrB ejecta will produce as they undergo flash irradiation. We are unaware of any previous published predictions of expected locations of light echoes for any astronomical object other than SN 1987A \citep{schaefer1987}.

In Section 2 we describe the observational data of this paper. The images  of the newly discovered nebulosity surrounding T CrB are presented in Section 3. The ejecta energetics and predicted light echoes are discussed in section 4. We briefly summarize our results in Section 5. 

\section{The Data}\label{sec:datasets}

The imaging data of this study were gathered with the Condor Array Telescope \citep{Lanzetta2023}. Condor is comprised of six fast (f/5) apochromatic refractors with 180 mm objectives. Each refractor is equipped with a 9576 x 6388-pixel CMOS camera covering 2.3 x 1.5 $\mathrm{deg^{2}}$, with a plate scale of 0.85 arcsec/pixel. Details of the telescope, and image-signature removal (flat-fielding etc.), and astrometry are given in \citet{Lanzetta2023}.

T CrB was imaged with Condor through four narrowband filters, each of 3 nm FWHM, with the following central wavelengths (CWL): [OIII] 500.7 nm, H$\alpha$ 656.3 nm, [NII] 658.4 nm and [SII] 673.2 nm. In addition it was imaged through a broadband Luminance filter with a FWHM of 280 nm and a CWL of 550 nm. 

The telescope was dithered by a random offset of $\sim$ 15 arcmin between each 
exposure. Because different filters were used on different telescopes at different 
times, we define the “reach” of an observation obtained by Condor observations as 
the product of the total objective area and the total exposure time devoted to the 
observation. As Condor consists of six individual telescopes, each of objective 
area 0.0254 $m^{2}$, a one-second exposure with one telescope of the array yields a reach 
of 0.0254 $m^{2}$ s, and a one-second exposure with the entire array (i.e. with all six 
telescopes) yields a reach of 6 x 0.0254 $m^{2}$ s = 0.153 $m^{2}$ s. The dates of observation and reaches of the images of T CrB are presented in Table 1. Three or four telescopes were often used simultaneously with H$\alpha$ filters, accounting for the deeper reach in that filter.

\begin{table}[ht]
\centering
\begin{tabular}{p{1.00in}cccc}
\multicolumn{4}{c}{{\bf Table 1:}  Details of Observations} \\

\multicolumn{1}{c}{Condor Field 155930+255513} &\multicolumn{1}{c}{J2000} & \multicolumn{1}{c}{R.A. = 15:59:30} &\multicolumn{1}{c}{Dec = +25:55:13}\\
\hline
\hline

\multicolumn{1}{l}{Filter} & \multicolumn{1}{c}{Start Date} & \multicolumn{1}{c}{End Date} &\multicolumn{1}{c}{Reach ($m^{2}$s)} \\
\hline
\multicolumn{1}{l}{Luminance 550.0 nm}  & 2023-09-21 & 2024-05-31 & 3,389.5 \\
\multicolumn{1}{l}{[O III] 500.7 nm}    & 2023-09-03 & 2024-05-29 & 2,168.1 \\
\multicolumn{1}{l}{H$\alpha$ 656.3 nm}   & 2023-08-30 & 2024-05-30 & 10,077.0  \\
\multicolumn{1}{l} {[N II] 658.4 nm}     & 2024-03-22 & 2024-05-30 & 1,190.9 \\
\multicolumn{1}{l} {[S II] 671.6 nm}     & 2023-08-30 & 2024-05-28 & 4,519.4 \\

\hline
\end{tabular}
\end{table}

\section{T C\MakeLowercase{r}B Images}\label{}
\subsection{Narrowband and Luminance Images}

The four narrowband images of T CrB, taken through the filters noted above, as well as the broadband Luminance filter image, are shown in Figure 1. In Figure 2, we show the same images after applying a ``block-summing" procedure, which sums each 8x8 block of pixels in the original image to produce one pixel in the new image. While no obvious nebulosities are seen in Figure 1, faint nebulous structures {\it are} seen near T CrB in three of the four narrowband images of Figure 2: H$\alpha$, [NII], and [SII].

The brightest areas in the H$\alpha$ image are radiating with a surface brightness $\sim$ 6.6 x $10^{-18}\,\mathrm{erg/s/cm^{2}/arcsec^{2}}$. The monochromatic 1-sigma uncertainty per pixel of the H$\alpha$ image is $\sim$ 0.43 microJy, which corresponds to an energy flux 1-sigma uncertainty per pixel of 3.0 x $10^{-18}$\,erg/s/$\mathrm{cm^{2}}$; a surface brightness 1-sigma uncertainty over one pixel of 4.1 x $10^{-18}\,\mathrm{erg/s/cm^{2}/arcsec^{2}}$; and a surface brightness 3-sigma uncertainty over a 10 x 10 $\mathrm{arcsec^{2}}$ region of 4.1 x $10^{-18}$\, $\mathrm{erg/s/cm^{2}/arcsec^{2}}$.

In Figure 3 we show the four difference images of the area surrounding T CrB (Luminance minus each of [OIII], H$\alpha$, [NII] and [SII]). While the nebulosities seen are broadly similar in morphology, they differ significantly from each other in detail. The presence of strong [NII] emission, but none in [OIII], suggest that the nebulosities are shock-energized rather than photoionized. The nebulosity appears concentrated in two lobes, whose sketched outlines are superposed on the narrowband [SII] difference image. 

In Figure 4 we show a color composite which is a sum of the three difference images of the Luminance image minus H$\alpha$, [NII], and [SII], respectively. The bi-lobed structure is again visible. The reality of this structure will only be testable once spectroscopy is carried out to determine the dynamics of the ionized gas of this newly-detected NSR.  

In Figure 5 we show the ``star-cleaned" image of Figure 4, created with the V2 version of StarNet2 ({\url{https://www.starnetastro.com/experimental}}). 
In Figure 6, we show ``star-cleaned" versions of the individual channels of the RGB image, created with the same tool and convolved with a 2D Gaussian kernel ($\sigma_x = \sigma_y = 5$) to highlight the structure of the nebulosity. The 2-degree sized, bi-lobed structure suggested by Figure 3 remains evident.

\newpage
\section{Implications of the Images}

\subsection{Could the Nebulosity NOT be due to an NSR?}

NSRs are {\it predicted} to exist around all RN \citep{Darnley2019, Healy-Kalesh2023} as the inevitable consequence of i) velocity gradients in nova ejecta, and ii) recurrence times sufficiently short that the early, high-velocity ejecta of RNe will eventually catch up to, collide and merge with the slow-velocity ejecta of the previous nova. This reasoning prompted us to use Condor to deeply image the region around the RN-candidate KT Eri \citep{brad2022}, revealing its H$\alpha$-bright NSR \citep{Shara2024}, and confirming its RN nature. While degree-sized H$\alpha$ structures are not expected at the relatively high Galactic latitude ($\sim$ -32 deg) of KT Eri, and the nebulosity surrounding KT Eri was {\it predicted}, it was nonetheless Southern African Large Telescope (SALT) spectra that {\it proved} that the nebulosity surrounding KT Eri both matched it in radial velocity, and possessed the modest velocity dispersion of a NSR. 

 The even more extreme Galactic latitude of T CrB (+48 deg) suggests that the chance superposition of a $\sim$ 2 degree-sized nebulosity with the RN is very unexpected a priori. The lack of [OIII] emission in Figures 3 and 4 rules out the possibility that the nebulosity is a planetary nebula or a supernova remnant \citep{Winkler2021,Long1990}. Still, despite the {\it prediction} that T CrB should be surrounded by a NSR, our discovery of the nebula of Figures 2-6 does not prove, with certainty, that the nebula really is a NSR and not a line-of-sight HII region. 
 
 We therefore searched the Gaia DR3 database to determine whether we could find a star within the Condor FOV that is both sufficiently hot (at least $\sim$ 30 kK) and sufficiently luminous (at least $\sim\, 100 L_{\odot}$) to ionize the hydrogen in its vicinity, along our line of sight to T CrB. Our search criteria were that the Gaia absolute G magnitude be $<\,$ 0.0; that the BP - RP color be $<\,$ -0.2; and that the apparent G-band magnitude be $<\,$ 10.0. Not a single star in the Condor FOV meets these three criteria simultaneously. The closest match is the star Gaia DR3 1220093045066977408 = BD +26 2766, with a Gaia parallax of 0.5499 $\pm$ 0.0302 mas, placing it at $\sim$ twice the distance of T CrB. Its absolute G mag = -0.41, its color is BP-RP = -0.13, and its apparent G mag = 10.89. The Gaia Dr3 catalog lists its temperature as 15.1 kK, while SIMBAD returns a spectral type of A0. This star is far too cool to ionize hydrogen; it cannot be responsible for the nebulosity of Figures 2-6. Absent any star in the Condor FOV that is both luminous and hot enough to ionize hydrogen, we conclude that the nebulosity surrounding T CrB is {\it not} an HII region. We predict that spectroscopy at multiple points throughout the nebula, likely with an 8-meter class telescope, will ascertain a match between the radial velocity of our claimed NSR relative to that of T CrB. It will also show an expansion velocity of order tens of km/s, typical of NSRs.

\subsection{The T CrB NSR Age}

The components of T CrB's Gaia-measured proper motion $\mu$ are $\mu_{\alpha}$ = -4.461 $\pm$ 0.016 mas/yr and $\mu_{\delta}$ = 12.016 $\pm$ 0.028 mas/yr \citep{Gaia2023}. The white arrow in Figure 4 corresponds to 100,000 yr of proper motion of T CrB, which is 1281". This is $\sim$ 1/2 the radius of the nebulosity of Figs. 4 and 5, suggesting that the NSR (nearly stationary) ejecta are at least $\sim$ 200,000 yrs old. It is possible that even Condor's remarkably large field-of-view has not captured the full extent of T CrB's ejecta, so that its NSR might be considerably larger (and older) than the above simplistic estimate. 

\subsection{The T CrB NSR size and mass}

The largest nova shells of 20th century novae are $\sim$ 1 pc in size or smaller. In contrast, NSR sizes are ultimately limited by the interstellar medium material swept up by the many thousands of RN eruptions' ejecta. The total ejected mass of a RN cannot greatly exceed $\sim$ 1 $M_{\odot}$, i.e. the mass of the donor star. In practice this limits NSR \citep{Healy-Kalesh2023} in the Galactic plane to $\sim$ 10 - 100 pc in size and $\sim 10^{3} - 10^{6} M_{\odot}$ in mass. Achieving this size brings the $\sim$ 1 $M_{\odot}$ of ejecta of a RN ``nearly" to a halt (relative to the initial 5,000 - 10,000 km/s ejection velocities of RN), with terminal velocities of just $\sim$ 10 - 100 km/s \citep{Healy-Kalesh2023}. This limits any further, substantial NSR size growth before the mass donor's envelope is depleted. 

As T CrB is located at a Gaia-determined distance of\, 914 $\pm$ 23 pc \citep{schaefer2022}, the $\sim$ 2 degrees in angular scale of the NSR seen in Figures 2-6 corresponds to 31.9 $\pm$ 0.8 pc. While this is smaller than the $\sim$ 50 pc-sized NSR of KT Eri \citep{Shara2024}, and the $\sim$ 134 pc-sized NSR of M31-12a \citep{Darnley2019}, it well within the range in sizes of NSR predicted by the hydrodynamical models of \citet{Healy-Kalesh2023}, especially ``younger" NSR with ages of order just a few x $10^{5}$ yr.

\subsection{The T CrB NSR Energetics}

As already noted, the brightest areas in the H$\alpha$ image are radiating with a surface brightness $\sim$ 6.6 x $10^{-18}\,\mathrm{erg/s/cm^{2}/arcsec^{2}}$. Assuming that the ejecta are radiating isotropically, and at the 914 pc distance to T CrB also noted above, their H$\alpha$ luminosity is of order 5 L$_{\odot}$. Other hydrogen lines, particularly of the Lyman series, may exceeed that value, but only by factors of order a few. The energy required to maintain the observed NSR hydrogen-line luminosity for $\sim$ 200,000 yr is thus $\sim$ a few times 10$^{47}$ erg. 

The mass transfer rate required for a RN eruption every $\sim$ 80 yr from a 1.37 M$_{\odot}$ white dwarf is $\simeq$ 1 - 3 x $10^{-8}$ M$_{\odot}$/yr, and about half that accreted mass ($\sim$ $10^{-6}$ M$_{\odot}$) will be ejected \citep{Hillman2016} at velocities $\sim$ 3000 km/s \citep{mcl1946}. The kinetic energy of the mass ejected during each eruption is thus $\sim$ $10^{44}$ erg, so the $\sim$ 2500 eruptions of the past 200,000 yr have injected $\sim$ 2.5 x $10^{47}$ erg of kinetic energy into the NSR, in good agreement with the line luminosity estimate of the previous paragraph. Below we show that while the NSR is irradiated by a RN flash every $\sim$ 80 yr, it is so optically thin that the photons emitted by the nova eruption do not measurably increase its energy budget. We conclude that the observed luminosity of T CrB's NSR is accounted for by the deposition of the kinetic energy of the nova ejecta alone.
\subsection{Light Echoes}

\subsubsection{The Shells inside the NSR}

Because the white dwarf of T CrB is expected to undergo a thermonuclear-powered RN eruption during the coming one to two years \citep{brad2024,Schneider2024}, a flash of nova light comprising $\sim 10^{44} \mathrm{erg}$ will propagate outwards through the large nebulosities reported in this paper, at the angular  rate of $\sim$ 50 pixels/yr. As already noted, the lack of [OIII] emission and presence of [NII] emission (see Figure 3) suggests that the nebulosities are currently shock-ionized rather than photoionized. Might this change, especially in the inner parts of the nebula, during the first years after T CrB's coming eruption? In addition, could currently ``dark" material (similar to that detected around the RN T Pyx) also light up \citep{Shara1997,shara2015}? If so, these could enable a full 3D model of the T CrB ejecta. These would be just the second set of light echoes whose morphology was ever {\it predicted} before the transient event that gave rise to them (see Schaefer 1987 for predictions of SN 1987A's light echoes). Unfortunately, the following modeling suggests that we will see no fluorescent light echoes, though continuum light echoes might be observed.

\subsubsection{A Model of Fluorescent Light Echoes of Recent T CrB Ejecta}

We assume that $10^{-6} M_{\odot}$ of hydrogen is ejected from the white dwarf of T CrB during every eruption, and that all eruptions are spaced exactly 80 years apart. We also assume that the rate of ejection is zero outside of eruption, and uniform for $10^{6}$ s during eruption; that all matter is ejected at 3,000 km/s \citep{mcl1946}, i.e. 1\% the speed of light; and that all matter is ejected with spherical symmetry. We also assume that the radius to which the ISM surrounding T CrB has been swept is 15 pc (the observed radius of its NSR) so that all recent ejections of mass are into a near vacuum; and that, on the basis of the enhanced brightness to the SE of T CrB that is seen in Figure 5, the thickness of the NSR shell is $\sim$ 10\% of its radius, i.e. 1.5 pc. so that its mass-filled volume is $\sim$ 4 x $10^{59} \mathrm{cm^{3}}$. The NSR simulations of \citep{Healy-Kalesh2023} support these latter assumptions, especially the generation of a near-vacuum environment around a RN by the sweeping of most circumstellar matter by thousands of successive mass ejections. 

Under these assumptions, each successive ejected shell coasts independently until it runs into the ``pileup" of swept-up ISM located $\sim$15 pc from, and surrounding T CrB. Moving at 1\% the speed of light, it takes each ejected shell nearly 5,000 yr to reach the NSR, so we expect $\sim$ 60 such nested shells to occupy the volume between T CrB and its NSR.  

We first consider the effects of the next RN flash of T CrB on the ejecta of 1946. Moving at 3,000 km/s, those ejecta are currently a shell with a wall thickness of $\sim$ 3 x $10^{14}$ cm and a shell radius of $\sim$ 7.2 x $10^{17}$ cm. The density of the $10^{-6} M_{\odot}$ of mass within the shell will be of order $10^{-23} \mathrm{gm/cm^{3}}$ or $\sim$ 5 atoms/$\mathrm{cm^{3}}$. The optical depth (opacity x density x path-length) of that shell is of order $10^{-9}$, so that just $\sim$ $10^{35}$ erg of the $\sim$ $10^{44}$ erg flash will be absorbed by the shell. This is only sufficient energy to increase the fraction of ionized hydrogen in the shell by $\sim$ $10^{-6}$.

The e-folding time over which electrons recombine in ionized hydrogen is $10^{5}$/$\mathrm{N_{e}}$ yr, where $\mathrm{N_{e}}$ is the number density of free electrons \citep{osterbrock1989}. The recombination time of ionized hydrogen in the 1946 shell will be $\sim$ $10^{5}$/5 yr = 20,000 years immediately after the flash energy is absorbed, and monotonically increasing as the shell continues to expand. The rate of emission of the newly absorbed flash energy will be $\sim$ $10^{35}$ erg/ 20,000 yr $\sim$ $10^{-9} L_{\odot}$. This is $\sim$ 14 orders of magnitude fainter than T CrB near its peak i.e. $\sim$ 36th magnitude if the shell were an unresolved point source. In fact the 1946 shell is $\sim$ 0.5 pc in diameter, which will be seen as a $\sim$ 113" diameter sphere covering $\sim$ 10,000 $\mathrm{arcsec^{2}}$. The surface brightness of the fully illuminated 1946 shell will thus be an utterly undetectable $\sim$ 46 mag/$\mathrm{arcsec^{2}}$.  

The shells of 1866, 1786, and 1706 have twice, three times and four times the radius and hence one fourth, one ninth and one sixteenth the density and optical depth of the shell of 1946. The infinite series sum of $1/n^{2}$ is $\pi^{2}$/6 $\sim$ 1.64, so that the flash energy absorbed by all $\sim$ 60 shells surrounding the 1946 shell will be only $\sim$ 64\% of that absorbed by the 1946 shell. Each of these shells will thus be even fainter than the undetectable 1946 shell.

\subsubsection{Light Echoes from the NSR itself}

An estimate of the mass and density of the swept-up ISM that comprises T CrB's NSR is needed in order to predict the strength of its light echoes. T CrB is located $\sim$ 680 pc above the Galactic Plane, where the ambient ISM density is likely to be an order of magnitude lower ($\sim$ 0.1 atoms/$\mathrm{cm^{3}}$) than than in the plane of the Galaxy. Assuming ambient densities of 0.1 or 1 atom /$\mathrm{cm^{3}}$, and a mass-filled volume $\sim$ 4 x $10^{59} \mathrm{cm^{3}} $, then the mass within the NSR shell is just $\sim$ 30 or 300 $M_{\odot}$, respectively. That mass will have an optical depth $<10^{-5}$, so just like the $\sim$ 60 shells expanding towards it, the NSR itself will not brighten enough due to absorption of photons from each RN flash to yield detectable light echoes.

Under the discrete shell model of section 4.5.2, a $\sim 10^{-6} M_{\odot}$ shell ejected $\sim$ 5,000 yr ago reaches the NSR shock pileup front every $\sim$ 80 yr, and is decelerated and assimilated into the NSR. This could generate NSR light echoes (even if the energy deposition is not via photons) if the assumptions of the simple nova ejection model of section 4.5.2 were strictly correct. But they are not. All novae are both observed and predicted by numerical simulations to eject matter at a wide range of speeds, with the fastest ejected material moving at 5-10X the speed of the slowest ejecta \citep{Warner1995}. Thus the discrete, thin shelled-structures we envisioned to roughly estimate the masses and densities of matter between T CrB and its NSR don't maintain their thin-walled shapes, or even identities for long. Within one or two inter-eruption times successive ejected shells overlap and mix, and by the time the ejecta reach the NSR shock boundary they should be a continuous and thoroughly mixed wind. This has no effect on the prediction of the previous section, namely that the still-coasting ejecta from T CrB are essentially transparent. In summary, we predict that there will be no detectable fluorescent light echoes in the close-in environs of T CrB or its extended NSR on any timescale.

\subsection{Dust and Continuum Light Echoes}

The declining light curve of the 1946 eruption of T CrB shows no sign of a sudden dip in brightness which would be expected if substantial amounts of dust formed in the ejecta. There was, however, a very significant dip (up to $\sim$ 1.5 mag) in the brightness of T CrB during 1945.1 through 1946.1    \citep{brad2023b}, and again beginning in March 2023. These brightness dips are simplest to interpret as dust produced by the red giant in the $\sim$ yr before a RN outburst. But there is no straightforward way to robustly turn these observed brightness dips (if they really are dust-induced) into estimates of dust mass, composition, grain sizes, geometry and dynamics. 

Lindblad (http://www.cbat.eps.harvard.edu/IAUCs/IAUC1030a.png) predicted the expected sizes of light echo rings, after an eruption, from the near and far sides of T CrB. These were based on an assumed distance to the nova of 614 pc and a spherically symmetric dust ejection velocity of 1000 km/s. Correcting to the Gaia-determined distance of 914 pc, and adopting 1000 km/s as the ejection velocity, nearside rings of expanding radii of 5", 10", 20" and 27.8" should be seen 2.4, 9.8, 45 and 149 d after the eruption begins. Contracting rings of radii 20", 10" and 5", from the remote side of the ejecta, should be seen 249, 285 and 292 d after the eruption. Higher velocity dust ejecta will proportionally shorten these timescales.

We conclude that dust-scattered, continuum light echoes may be detected around T CrB, but that insufficient input information is available to make an informed prediction about the intensities or morphologies of such echoes. {\it Hubble Space Telescope} optical imaging as well as {\it James Webb Space Telescope} infrared imaging, on timescales of days to at least one year, would be very worthwhile as probes of any such dust formation and ejection.

\section{Conclusions}

T CrB is now the third RN shown to be surrounded by an optically detected NSR. The NSR is at least $\sim$ 30 pc in extent and at least $\sim$ 200,000 yrs old. The ejecta of the three currently known NSR are all tens of parsecs in size, and are seen in narrowband H$\alpha$, [NII], and [SII] images, but not in [OIII] or broadband Luminance images. The morphologies of the gas of the line-emitting ions of T CrB are broadly similar, but differ in detail in the three filters due to varying shock conditions across the ejecta. Block-summed difference images suggest that the ejecta of T CrB might be bi-lobed. 

The ejecta of past RN eruptions will be irradiated at successively later times after T CrB undergoes its once-in $\sim$ 80-year eruption in the near future. We simplistically modeled $\sim$ 60 concentric, thin shells of $\sim$ $10^{-6} M_{\odot}$ expanding uniformly away from T CrB, ejected in the past $\sim$ 5,000 years. These shells' low masses and very small optical depths imply that their surface brightnesses due to nova flash irradiation will be far too low to be detectable. This conclusion is unchanged under the more realistic assumption that the multiple ejections' matter blends into a wind within a few centuries. In similar vein, the NSR itself is so optically thin that its brightening due to the 1946 flash or the upcoming RN flash should be immeasurably small. 

A lack of dips in the post-eruption light curves of past T CrB eruptions means there is probably no significant dust produced in the expanding matter ejected from the white dwarf. But pre-eruption dips seen in 1945 and 2023 light curves suggest that the red giant of T CrB may have emitted an undetermined amount of dust with an unknown geometry and density. We therefore cannot predict the morphology or intensity of continuum light echoes due to dust scattering following the coming eruption. A simple model predicts the appearance and disappearance of continuum light echoes, from dust produced in 1946, on timescales of days to $\sim$ 1 yr.

In summary, we conclude that i) there will be no fluorescent light echoes seen after T CrB erupts, and ii) there might be dust-scattered, continuum light echoes as large as 20-30" in radius during the year following the nova eruption, but we cannot predict their intensity or morphology. We also predict that iii) that the ongoing NSR power source is the transformation of the kinetic energy of merged, ejected shells into optical light via collisions which continuously ionize the $\sim$ 30 - 300 $M_{\odot}$ of swept-up ISM surrounding T CrB. 

\section{Acknowledgments}
KML, MMS and JTG acknowledge the support of National Science Foundation Grants 1910001, 2107954, 2108234, 2407763, and 2407764. JM acknowledges support from the Polish National Science Center grant 2019/35/B/ST9/03944. This work has made use of data from the European Space Agency (ESA) mission {\it Gaia} (\url{https://www.cosmos.esa.int/gaia}), processed by the {\it Gaia} Data Processing and Analysis Consortium (DPAC,\url{https://www.cosmos.esa.int/web/gaia/dpac/consortium}). Funding for the DPAC has been provided by national institutions, in particular the institutions participating in the {\it Gaia} Multilateral Agreement. We thank Brad Schaefer and the anonymous referee for excellent suggestions which helped improve the paper.

\bibliography{TCrB}
\bibliographystyle{aasjournal}

\newpage
\begin{figure*}[h!]
    \centering
    \includegraphics[width=7in]{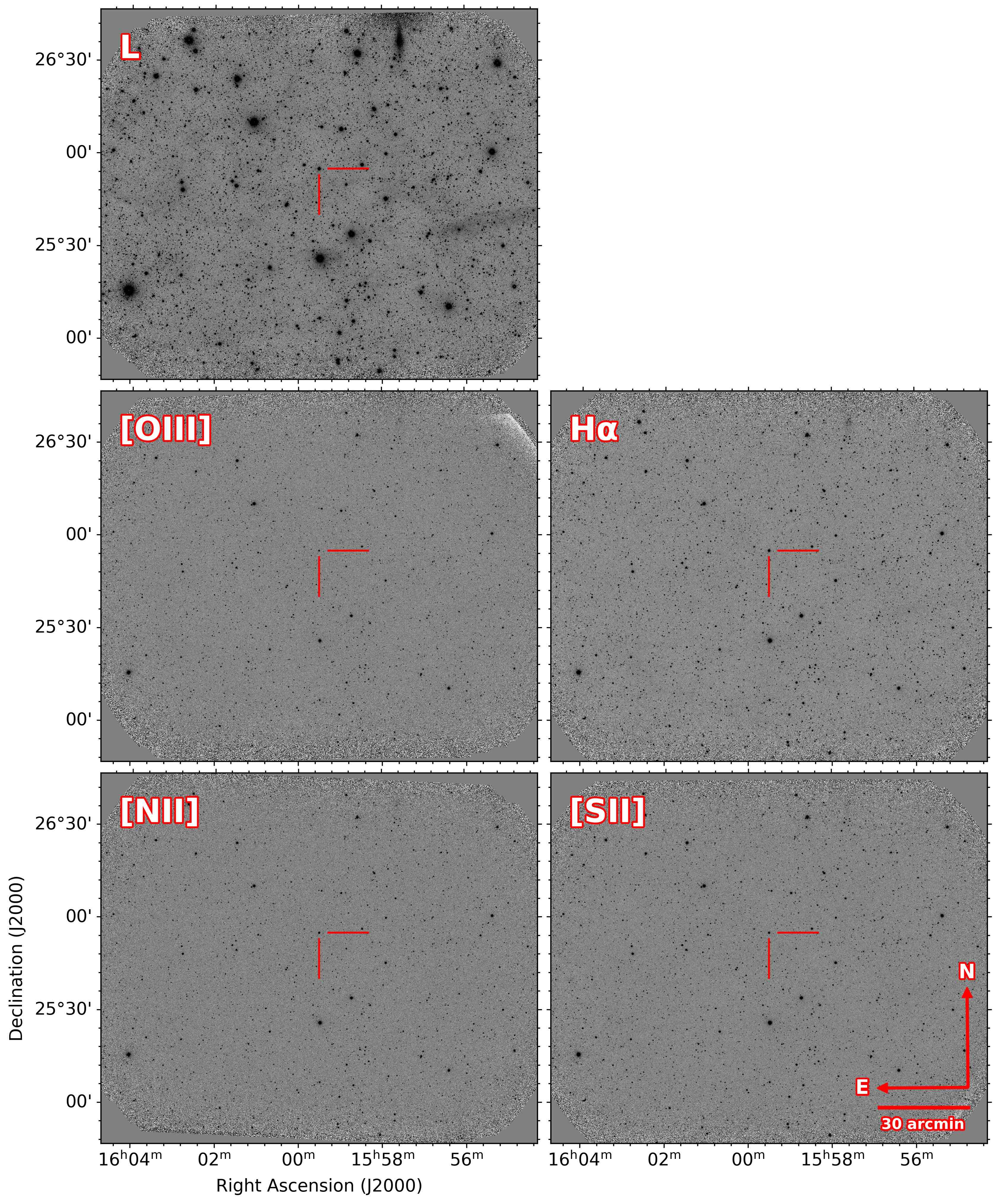}
    \caption{Co-added images of T CrB (position indicated with red ticks) in the Luminance broadband and four narrowband ([OIII], H$\alpha$, [NII], and [SII]) filters. Each image has a different contrast scaling corresponding to the parameters assigned by the ``zscale" function in SAO DS9.}
\end{figure*}

\newpage

\begin{figure*}[h!]
    \centering
    \includegraphics[width=7in]{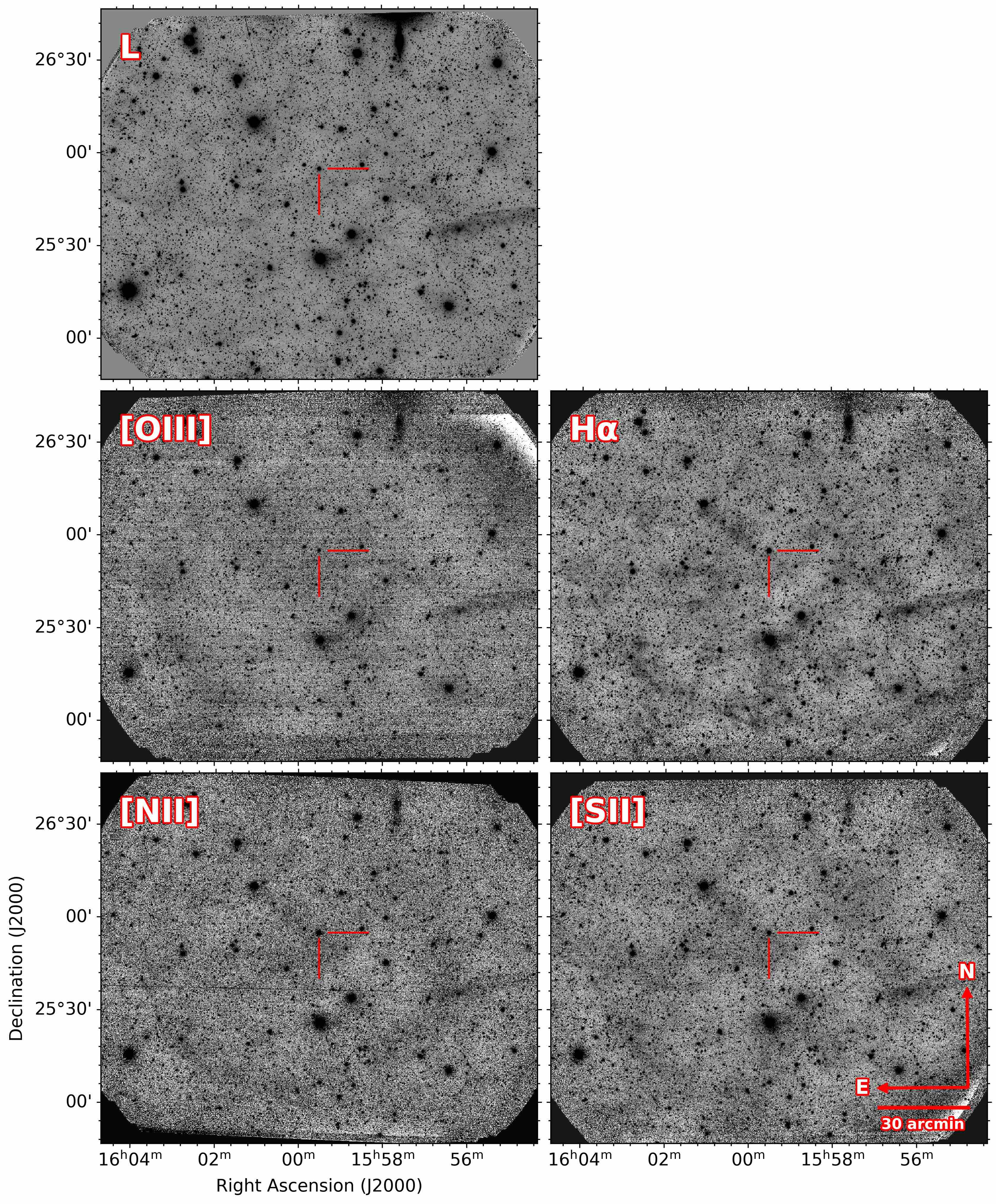}
    \caption{Same as Figure 1, but each image has been 8x8 block-summed (see text). Faint, nebulous structures are seen to surround T CrB in H$\alpha$, [NII] and [SII] in all directions.}
\end{figure*}

\newpage

\begin{figure*}[h!]
    \centering
    \includegraphics[width=7in]{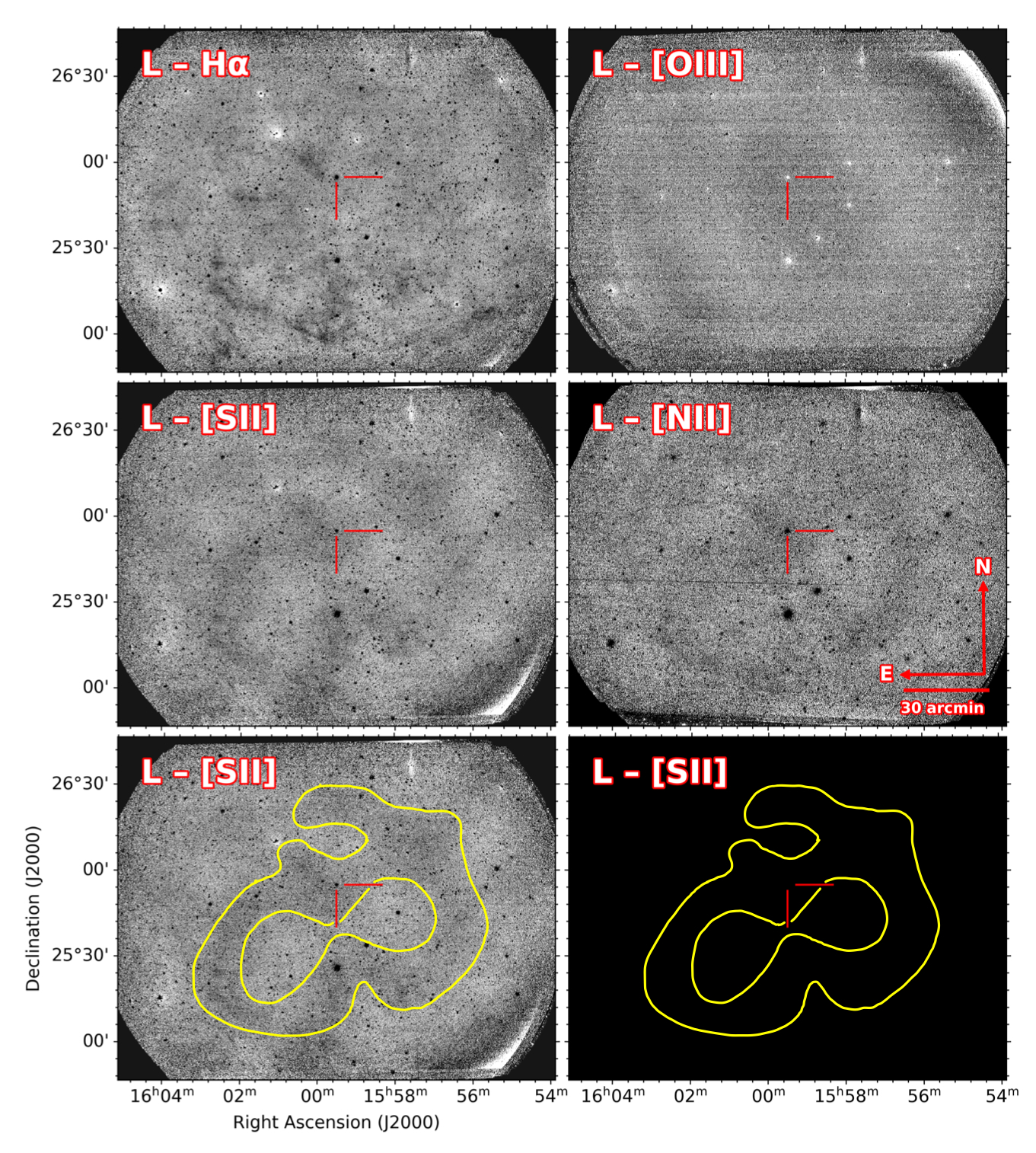}
    \caption{Difference images of T CrB (position indicated with red ticks) created by directly subtracting each of the 4 narrowband coadded images from the Luminance coadded image, then applying 8x8 block-summing as in Figure 2. A hand-sketch of the two possible lobes of the nebulosity are drawn over a copy of the [SII] narrowband difference image. All images have the same scaling to allow for direct comparison of the faint emission features. The presence of clear emission in [NII], but none in [OIII], demonstrates that the nebulosities are shock-ionized rather than photoionized.}
\end{figure*}

\newpage

\begin{figure*}[h!]
    \centering
    \includegraphics[width=7in]{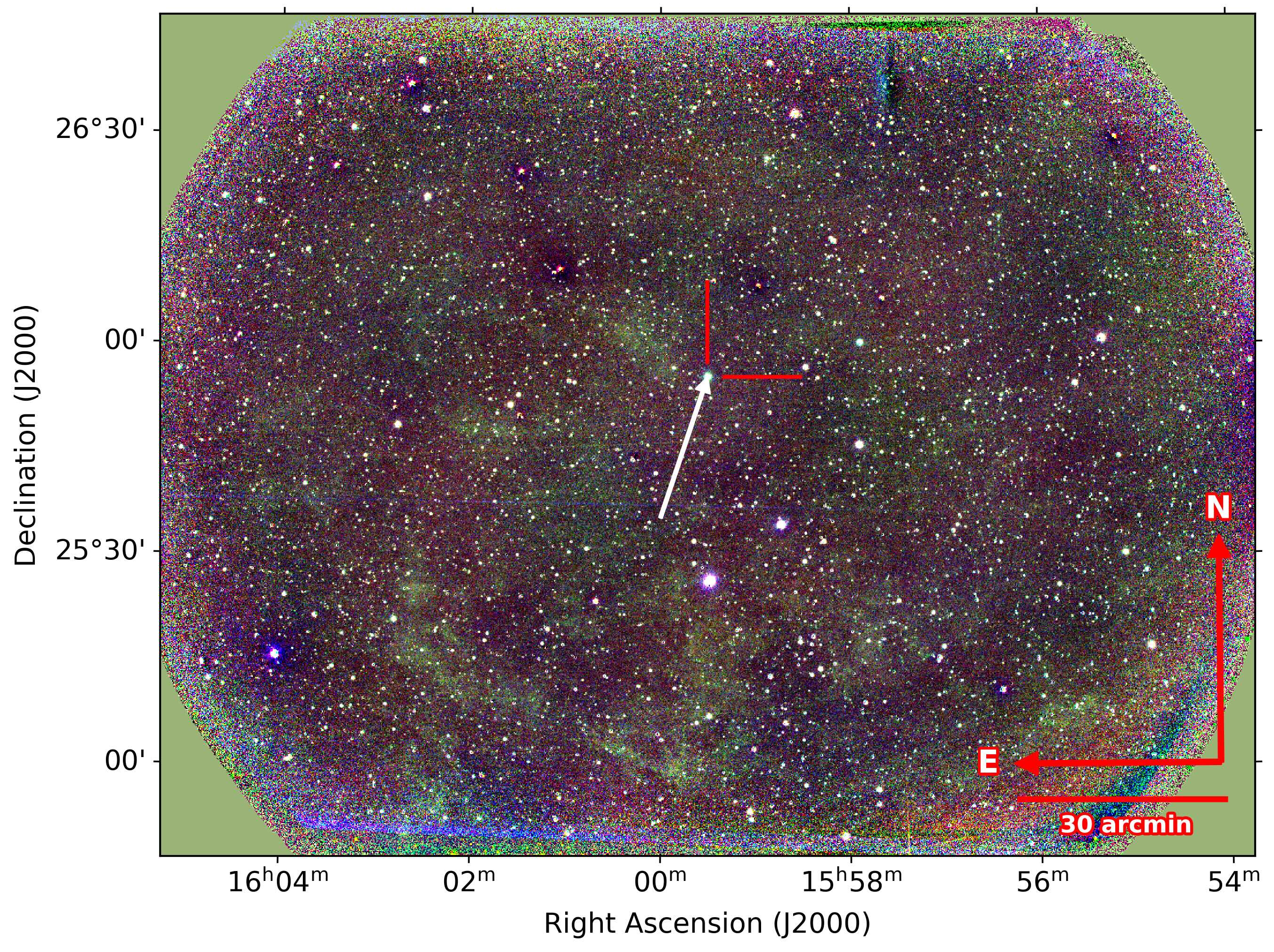}
    \caption{RGB image of T CrB (position indicated with red ticks) and its surrounding emission. Each of the three channels is a block-summed difference image (see Figure 3). Red corresponds to Luminance - [SII]; green corresponds to Luminance - H$\alpha$; and blue corresponds to Luminance - [NII]. The intensity scaling of each channel has been adjusted individually to highlight the emission features. There is a suggestion of a bi-lobed structure (also noted and sketched in Figure 3). The white arrow points in the direction of the star's Gaia-measured proper motion, with a length corresponding to the angular distance that T CrB has moved in the past 100,000 yrs.}
\end{figure*}

\newpage

\begin{figure*}[h!]
    \centering
    \includegraphics[width=7in]{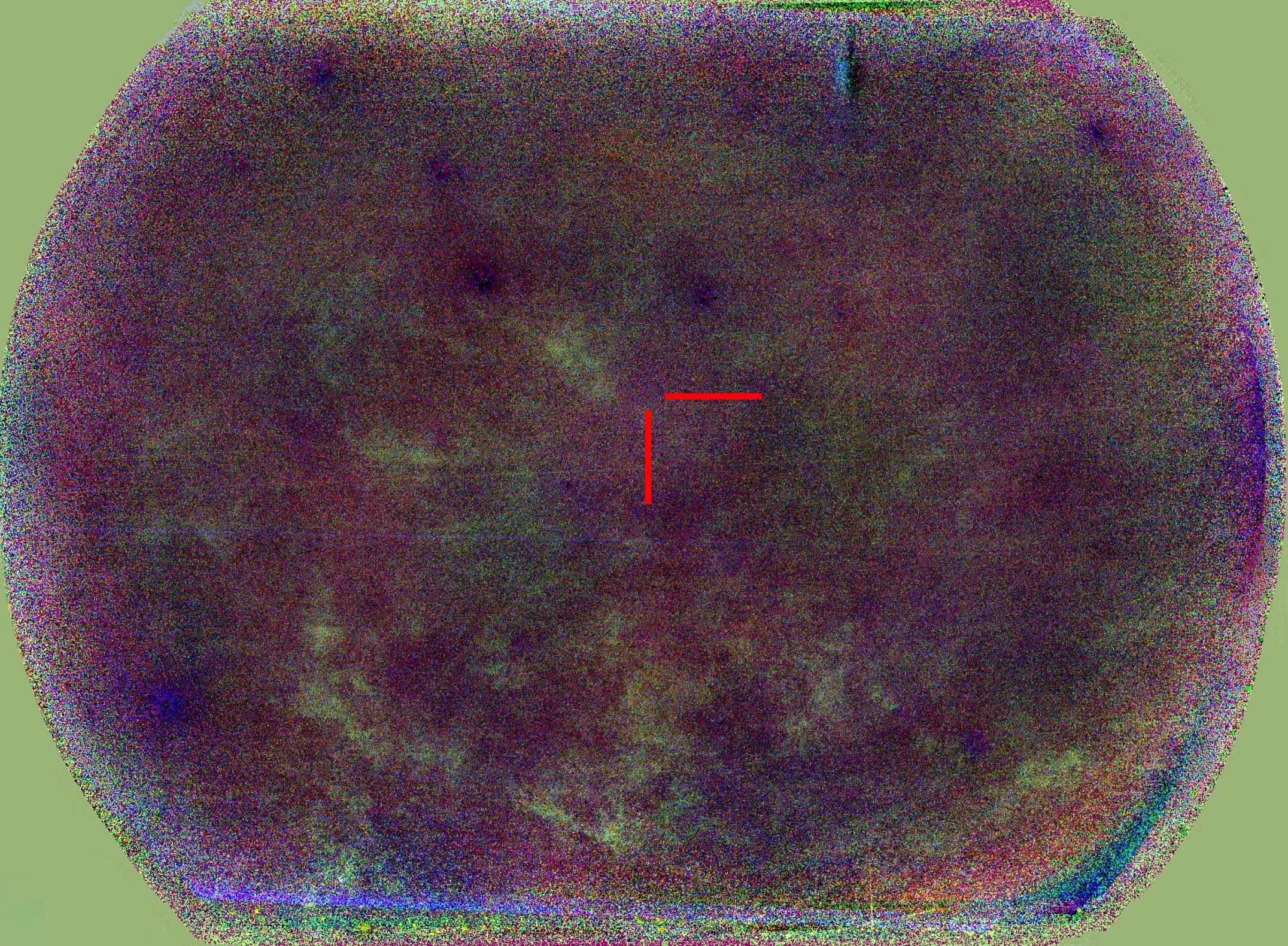}
    \caption{Same as Figure 4, but after the ``star-cleaning" algorithm STARNET2 is applied. The bi-lobed structure is again seen, with a maximum diameter of $\sim$ 2 deg in the NW to SE direction. The location of T CrB is indicated with red ticks.}
\end{figure*}

\newpage

\begin{figure*}[h!]
    \centering
    \includegraphics[width=3.5in]{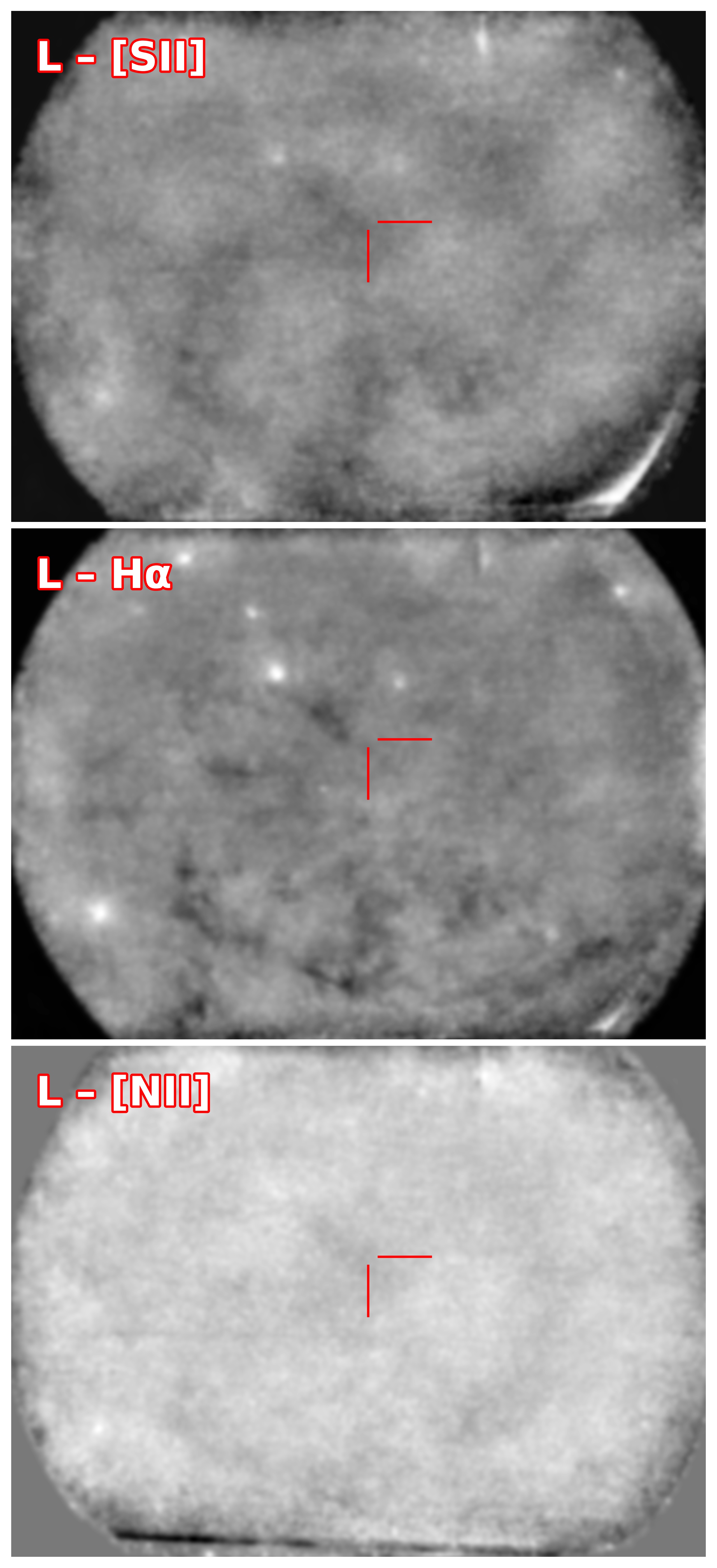}
    \caption{``Star-cleaned" versions of the Luminance - [SII], Luminance - H$\alpha$, and Luminance - [NII] images, created with STARNET2. The scaling of each image matches the scaling of the channels in the RGB image (Figures 4 and 5). The images have been smoothed with a 2D Gaussian kernel to emphasize the nebulosity. Note that the (noisy) [NII] image has the lowest reach (and highest background) of the three narrowband images shown. The position of T CrB is indicated with red ticks.}
\end{figure*}

\end{document}